\documentclass[twocolumn,aps,superscriptaddress,showpacs,floatfix]{revtex4}
\usepackage{amssymb}
\usepackage{amsmath}
\usepackage{graphicx}
\usepackage[normalem]{ulem}
\usepackage[dvips]{color}
\usepackage{bm}
\usepackage{longtable}

\renewcommand\sout{\bgroup \color{red} \ULdepth=-.5ex \ULset}

\begin{document}

\title{Neutron-proton effective mass splitting in neutron-rich matter at normal density from analyzing nucleon-nucleus scattering data within an isospin dependent optical model}

\author{Xiao-Hua Li}
\affiliation{Department of Physics and Astronomy, Texas A$\&$M University-Commerce, Commerce, TX 75429-3011, USA}
\author{Wen-Jun Guo}
\affiliation{College of Science, University of Shanghai for Science and Technology, Shanghai, 200093, China}
\affiliation{Department of Physics and Astronomy, Texas A$\&$M University-Commerce, Commerce, TX 75429-3011, USA}
\author{Bao-An Li\footnote{%
Corresponding author: Bao-An.Li$@$tamuc.edu}}
\affiliation{Department of Physics and Astronomy, Texas A$\&$M University-Commerce, Commerce, TX 75429-3011, USA}
\affiliation{Department of Applied Physics, Xi'an Jiaotong University, Xian 710049, China}
\author{Lie-Wen Chen}
\affiliation{INPAC, Department of Physics and Astronomy and Shanghai Key Laboratory for Particle Physics and Cosmology, Shanghai Jiao Tong University, Shanghai 200240, China}
\affiliation{Center of Theoretical Nuclear Physics, National Laboratory of Heavy Ion Accelerator, Lanzhou 730000, China}
\author{Farrukh J. Fattoyev}
\affiliation{Department of Physics and Astronomy, Texas A$\&$M University-Commerce, Commerce, TX 75429-3011, USA}
\author{William G. Newton}
\affiliation{Department of Physics and Astronomy, Texas A$\&$M University-Commerce, Commerce, TX 75429-3011, USA}

%\date{\today}

\begin{abstract}

The neutron-proton effective mass splitting in asymmetric nucleonic matter of isospin asymmetry $\delta$ and normal density is found to be
$m^{*}_{n-p}\equiv(m^{*}_{n}-m^{*}_{p})/m=(0.41 \pm0.15)\delta$  from analyzing globally 1088 sets of reaction and angular differential cross sections of proton elastic scattering on 130 targets with beam energies from 0.783 MeV to 200 MeV, and 1161 sets of data of neutron elastic scattering on 104 targets with beam energies from 0.05 MeV to 200 MeV within an isospin dependent non-relativistic optical potential model. It sets a useful reference for testing model predictions on the momentum dependence of the nucleon isovector potential necessary for understanding novel structures and reactions of rare isotopes.
\end{abstract}

\maketitle

\section{Introduction}
Because of the finite range of the nuclear isovector interaction and the isospin dependence of Pauli blocking, the nucleon isovector (symmetry) potential in isospin-asymmetric nucleonic matter is momentum dependent, see, e.g., Refs. \cite{sjo76,Bom91,Dob99,LiBA08,zuo05,Fuchs1,Fuchs2}. Thus, neutrons and protons are expected to have different effective masses used to characterize the momentum dependence of their respective mean-field potentials in isospin-asymmetric nucleonic matter. Is the effective mass of neutrons larger, equal or smaller than that of protons in neutron-rich nucleonic matter? While it has significant ramifications on addressing many interesting issues in both nuclear physics and astrophysics, the theoretical answer to this question depends strongly on the model and interaction used~\cite{Bom01,Riz04,Li04}. For instance, among the 94 Skyrme interactions examined within the Skyrme-Hartree-Fock approach in Ref. \cite{OuLi11}, 48/29/17 of them predict a positive/negative/zero value for the neutron-proton effective mass splitting. One of the main reasons for this unfortunate situation is our poor knowledge about the in-medium properties of nuclear isovector interaction and the lack of reliable experimental probes of the neutron-proton effective mass splitting. Moreover, it is worth emphasizing that the neutron-proton effective mass splitting is simply part of the nuclear symmetry energy~\cite{XuC10,XuC11,Rchen} according to the Hugenholtz-Van Hove (HVH) theorem \cite{hug}. The symmetry energy encodes the energy related to the neutron$-$proton asymmetry in the equation of state of isospin asymmetric nuclear matter and is a key quantity for understanding many issues in nuclear physics and astrophysics~\cite{TEsym}. In fact, one of the major causes for the still poorly known density dependence of the nuclear symmetry energy is the uncertain momentum dependence of the isovector potential and the corresponding neutron-proton effective mass splitting~\cite{XuC10,LiBA13}. Therefore, it is imperative to reliably constrain the latter even at normal density. It is encouraging to note that some serious efforts have been made recently to find experimental observables sensitive to the neutron-proton effective mass splitting. For example, the single and/or double neutron/proton or triton/$^3$He ratio
at high momenta were found to be sensitive to the neutron-proton effective mass splitting consistently in several transport model studies of intermediate energy heavy-ion
collisions~\cite{LiBA04,Riz05,Feng12,Zhang14}. However, because of the simultaneous sensitivities of these observables to several not so well determined ingredients in transport models, no conclusion has been drawn from
heavy-ion collisions regarding the neutron-proton effective mass splitting yet. In principle, a more direct and clean approach of obtaining the neutron-proton effective mass splitting albeit only at normal density is using the energy/momentum and  isospin dependence of the nucleon optical potential from nucleon-nucleus scattering. Indeed, since the earlier 1960s, several parameterizations of the energy/momentum dependence of the nucleon isovector potential have been extracted using the data available at the time. However, these analyses are not completely independent and the parameterizations are valid in segmented energy ranges up to about 200 MeV. In Ref.~\cite{XuC10},
assuming that all of these nucleon isovector potentials are equally accurate and have the same predicting power beyond the original energy ranges in which they were analyzed, and by taking an average of the available 6 parameterizations a neutron-proton effective mass splitting of $(m_n^*-m_p^*)/m=(0.32\pm 0.15) \delta$ was obtained. Besides the rather rough assumption, we notice that the error bar was estimated by simply considering the range of the existing parameterizations of the optical potential which themselves do not have properly quantified uncertainties. In another recent attempt, using values of the symmetry energy and its density slope at normal density extracted from 28 different analyses of terrestrial nuclear laboratory data and astrophysical observations, $m^{*}_{n-p}=(0.27\pm0.25)\delta$ was extracted~\cite{LiBA13}. Here the error bar is a rough estimate as often the uncertainties of the individual entries for the analysis are not quantified. Thus, it is fair to state that currently there are clear experimental indications that the neutron effective mass is higher than that of protons at normal density. However, the exact value of the neutron-proton effective mass splitting has large uncertainties often not quantified. The situation at supra-saturation densities reached in heavy-ion collisions and/or the core of neutron stars is even worse.

The purpose of the present work is to provide a reliable value of the neutron-proton effective mass splitting at normal density with quantified uncertainty to be used as a reference to calibrate model predictions on the momentum dependence of nuclear symmetry potential in neutron-rich nucleonic matter. We achieve this goal by performing a global optical model analyses of all 2249 data sets of reaction and angular differential cross sections of neutron and proton scattering on 234 targets at beam energies from 0.05 to 200 MeV available in the EXFOR database at the Brookhaven National Laboratory~\cite{Exfor}. Moreover, the variances of all model parameters are evaluated consistently
by carrying out a covariance analysis of the error matrix around the optimized optical model parameters using the standard statistical technique detailed in Ref.~\cite{Rei10,Fat11}.  We found that the neutron-proton effective mass splitting is $m^{*}_{n-p}=(0.41\pm0.15)\delta$. To our best knowledge, this is currently the most stringent and reliable constraint on the neutron-proton effective mass splitting at normal density using a well established model from analyzing the complete data sets of the relatively simple nucleon-nucleus reactions.

The theoretical formalism and procedures we shall use are all well established in the relevant literature. For completeness and ease of discussions,  in Section II we shall first summarize the major ingredients of the non-relativistic isospin dependent optical potential model for nucleon-nucleus scattering.  After defining the neutron-proton effective mass splitting in terms of the momentum dependence of the isovector and isoscalar potentials in isospin-asymmetric nucleonic matter, we recall the general relationship between the nucleon optical potential in nucleon-nucleus scattering and the single-nucleon potential in nuclear medium. The results of our analyses are presented in Section III.
Finally, a summary is given in Section IV.

\section{Formalism}
In the following we outline the most important ingredients and the necessary steps for our extraction of the neutron-proton effective mass splitting at normal density from analyzing experimental data of nucleon-nucleus scattering up to the beam energy of approximately 200 MeV.

\subsection{Isospin dependent optical model for nucleon-nucleus scattering}
The optical model is a reliable tool for studying nucleon-nucleus scattering. For a historical review, we refer the reader to the textbook by  Hodgson~\cite{Hod94}. To access the available optical potentials for various applications, we recommend the reader to visit the section on optical models at IAEA's RIPL (Reference Input Parameter Library for theoretical calculations  of nuclear reactions) library \cite{Yon98}. Recent examples of developing local and/or global nucleon optical potentials from analyzing various sets of nucleon-nucleus scattering data can be found in
Refs.~\cite{Var91,Kon03,Wep09,Han10,LiXH12}. In this work, we restrict ourselves to nucleon-nucleus scattering below about 200 MeV where a non-relativistic description is appropriate~\cite{AnHX06}.

The phenomenological nucleon Optical Model Potential (OMP) for nucleon-nucleus scattering
$V(r,\mathcal{E})$ can be generally written as
\begin{eqnarray}\label{OMP1}
V(r,\mathcal{E})&=&-V_\mathrm{v} f_\mathrm{r}(r)-i W_\mathrm{v}
f_\mathrm{v}(r)+i 4 a_\mathrm{s}W_\mathrm{s}
\frac{\mathrm{d}f_\mathrm{s}(r)}{\mathrm{d}r}\nonumber\\
&+&2\lambda\!\!\!{^-}_\pi^2\frac{V_\mathrm{so}+iW_\mathrm{so}}{r}
\frac{\mathrm{d}f_\mathrm{so}(r)}{\mathrm{d}r}
\mathbf{S}\cdot\mathbf{L}+V_C(r)\,,
\end{eqnarray}
where the $V_\mathrm{v}$ and $V_\mathrm{so}$ are the depth of the real
parts of the central and spin-orbit potential, respectively; while the
$W_\mathrm{v}$, $W_\mathrm{s}$ and $W_\mathrm{so}$ are the depth
of the imaginary parts of the volume absorption, surface
absorption and spin-orbit potential, respectively; the $V_C(r)$ is the Coulomb potential for protons when they are used as  projectiles and is taken as the potential of a uniformly charged sphere with radius $R_C=r_C A^{-1/3}$, where $r_C$ is a parameter and $A$ is the mass number of targets. The $f_{i}$ ($i=\mathrm{r,v,s,so}$) are the standard Wood-Saxon shape
form factors; the $\mathcal{E}$ is the incident nucleon energy in the
laboratory frame; the $\lambda\!\!\!{^-}_\pi $ is the reduced Compton wave length
of pion and is taken as $\lambda\!\!\!{^-}_\pi= \sqrt{2.0}$ fm.

To more accurately extract useful information about the isospin dependence of the nucleon OMP, it is expanded to the second order in isospin asymmetry, i.e., $[(N-Z)/A]^2$ terms in the $V_\mathrm{v}$, $W_\mathrm{s}$ and $W_\mathrm{v}$. This term was found appreciable in two recent model studies and data analyses~\cite{Rchen,LiXH13}.
Moreover, the isoscalar part of $V_\mathrm{v}$ is expanded up to the quadratic  term in energy, i.e.,  $\mathcal{E}^2$. It is well known that this term is important to fit the nucleon-nucleus scattering data in both relativistic and non-relativistic descriptions~\cite{Hod94}. For the isospin-dependent parts, however, we found that the coefficient ratios of the second- to first-order terms in energy is about $10^{-5}$ to $10^{-3}$. To keep the number of parameters as small as possible, we neglect the quadratic terms in energy in the coefficients of the isospin dependent terms. Thus,  the following parameterizations for the $V_\mathrm{v}$, $W_\mathrm{s}$ and $W_\mathrm{v}$ are used in our  current analyses
\begin{align}
V_\mathrm{v}=&V_0+V_1\mathcal{E}
+V_2 \mathcal{E}^2+\tau_3(V_3+V_{3\mathrm{L}}\mathcal{E}) \frac{N-Z}{A}
\nonumber\\&+(V_4+V_{4\mathrm{L}}\mathcal{E})
\frac{(N-Z)^2}{A^2},\label{OMP2-1}\\
W_\mathrm{s}=&W_{\mathrm{s}0}+W_{\mathrm{s}1}\mathcal{E}+\tau_3(W_{\mathrm{s}2}+W_{\mathrm{s}2\mathrm{L}}\mathcal{E})
\frac{N-Z}{A}\nonumber\\
&+(W_{\mathrm{s}3}+W_{\mathrm{s}3\mathrm{L}}\mathcal{E})\frac{(N-Z)^2}{A^2},\\
 W_\mathrm{v}=&W_{\mathrm{v}0}+W_{\mathrm{v}1}\mathcal{E}
 +W_{\mathrm{v}2}\mathcal{E}^2\nonumber+\tau_3(W_{\mathrm{v3}}+W_{\mathrm{v3L}}\mathcal{E})\frac{N-Z}{A}\nonumber\\
&+(W_{\mathrm{v4}}+W_{\mathrm{v4L}}\mathcal{E})\frac{(N-Z)^2}{A^2},
\end{align}
where $\tau_3=+/-1$ for neutrons/protons. Denoting
the energy-dependent isoscalar potential
$\mathcal{U}_0(\mathcal{E})\equiv-(V_0+V_1 \mathcal{E}+V_2 \mathcal{E}^2)$,
the isovector (first-order symmetry) potential
$\mathcal{U}_{sym,1}(\mathcal{E})\equiv-(V_3+V_{\mathrm{3L}}\mathcal{E})$
and the second-order symmetry potential $\mathcal{U}_{sym,2}(\mathcal{E})\equiv
-(V_4+V_{\mathrm{4L}} \mathcal{E})$, the real part of the central potential
$\mathcal{U}_{\tau}(\mathcal{E})\equiv V_\mathrm{v}$ can be rewritten in the form
of the well-known Lane potential~\cite{Lan62},
\begin{align}\label{op2}
\mathcal{U}_{\tau}(\mathcal{E},\delta)=\mathcal{U}_0(\mathcal{E})+\tau_3\mathcal{U}_{sym,1}(\mathcal{E})\cdot\delta+\mathcal{U}_{sym,2}(\mathcal{E})\cdot\delta^2
\end{align}
where the isospin asymmetry $\delta$ is $(N-Z)/A$ for finite nuclei or $(\rho_n-\rho_p)/\rho$ for nuclear matter.  It is worth noting that the form factor peaks at the centers of target nuclei. Moreover, for medium and heavy nuclei, the central density is around the saturation density of nuclear matter. Thus, from nucleon scattering on medium to heavy targets, one can extract information about both the isoscalar and isovector potential and their energy dependences at the saturation density.

\subsection{Neutron-proton effective mass splitting and the momentum dependence of single-nucleon potential in isospin-asymmetric nucleonic matter}
Microscopic nuclear many-body theories indicate that the real part of the single-nucleon potential  $U_{\tau}(k, \mathcal{E},\rho,\delta)$ for $\tau=n$ or $p$ in nuclear matter of density $\rho$ and 
isospin-asymmetry $\delta$ depends on not only the nucleon momentum $k$ but also its energy $\mathcal{E}$, reflecting the nonlocality in both space and time of nuclear interactions, see, e.g. \cite{jamo,LLL}.  These two kinds of nonlocality can be characterized by using the so-called nucleon effective  k-mass and E-mass, respectively defined in terms of the partial derivative of $U$ with respect to $k$ and $\mathcal{E}$ \cite{jamo}. However, once a dispersion relation 
$k(\mathcal{E})$ or $\mathcal{E}(k)$ is known from the on-shell condition $\mathcal{E}=k^2/2m+U(k, \mathcal{E},\rho,\delta)$, an equivalent potential either local in space or time, i.e.,
$U(k(\mathcal{E}), \mathcal{E},\rho,\delta)$ or $U(k,\mathcal{E}(k),\rho,\delta)$, can be obtained.  The selection of a specific representation is often a matter of convenience in treating a given problem as the two expressions of $U$ are equivalent and easily transformable from one to the other. For example, while the phenomenological optical potential discussed in the previous section has been expressed as a function of energy only, it has long been well known that some parts of the energy dependence actually come from the explicit momentum dependence of $U$ due to the finite range of nuclear interaction. However, in the analyses of nucleon-nucleus scattering experiments within optical models, it is more convenient to use energy as a variable. The equivalent space-local potential $U(k(\mathcal{E}),\mathcal{E},\rho,\delta)$ is thus normally used in optical models. In this approach, while the momentum is not an independent variable explicitly, it not necessarily means that the potential is actually space-local completely. On the other hand, in some other applications, it is sometimes more convenient to use the equivalent time-local (static) potential $U(k,\mathcal{E}(k),\rho,\delta)$. For example, in transport model simulations of nuclear reactions one follows the evolution of nucleon phase space distribution function by solving Boltzmann-like equations using the $U(k,\mathcal{E}(k),\rho,\delta)$ as an input function. In this case, it is obviously more useful to express the potential as a function of momentum only.  Thus,  the nucleon effective mass can be calculated using either the first or second part of its defining equation depending on whether the $\mathcal{E}$ or $k$ is selected as the explicit variable \cite{jamo}
 \begin{eqnarray}\label{em1}
 \frac{m^{*}_{\tau}}{m_{\tau}}&=&1-\frac{dU_{\tau}(k(\mathcal{E}),\mathcal{E},\rho,\delta)}{d\mathcal{E}}\\ \nonumber
 &=&\left[1+\frac{m_{\tau}}{\hbar^2k_F^{\tau}}\frac{dU_{\tau}(k,\mathcal{E}(k),\rho,\delta)}{dk}\Bigg|_{k_F^{\tau}}\right]^{-1}
 \end{eqnarray}
where $m_{\tau}$ represents the mass of neutrons or protons in free-space and the neutron/proton Fermi momentum $k_F^{\tau}=(1+\tau_3\delta)^{1/3}\cdot k_F$ with $k_F=(3\pi^2\rho/2)^{1/3}$  being the nucleon Fermi momentum in symmetric matter at density $\rho$. 

Similar to the nucleon optical potential, the nucleon potential $U_{\tau}(k,\rho,\delta)$ in isospin-asymmetric matter can be written as
\begin{align}\label{sp}
U_{\tau}(k,\rho,\delta)&=U_0(k,\rho)+\tau_3 U_{sym,1}(k,\rho)\cdot\delta+U_{sym,2}(k,\rho)\cdot\delta^2\nonumber\\
&+\tau_3\mathcal{O}(\delta^3),
\end{align}
where $U_0(k,\rho)$, $U_{sym,1}(k,\rho)$ and $U_{sym,2}(k,\rho)$ are the isoscalar, isovector (first-order symmetry) and second-order symmetry potentials, respectively. 
The neutron-proton effective mass splitting
$m^*_{n-p}(\rho,\delta)\equiv(m_{\rm n}^*-m_{\rm p}^*)/m$ is then
\begin{align}\label{em2}
m^*_{n-p}=\frac{\frac{m}{\hbar^2}\left(\frac{1}{k_F^p}\frac{dU_p}{dk}\mid_{k_F^p}-\frac{1}{k_F^n}\frac{dU_n}{dk}\mid_{k_F^n}\right)}{\left[1+\frac{m_p}{\hbar^2k_F^p}\frac{dU_p}{dk}\mid_{k_F^p}\right]\left[1+\frac{m_n}{\hbar^2k_F^n}\frac{dU_n}{dk}\mid_{k_F^n}\right]}.
\end{align}
Since the $U_{sym}(\rho,k)\cdot\delta$ term is always much smaller than the isoscalar potential $U_0(\rho,k)$ in Eq.~(\ref{sp}), the denominator in Eq.~(\ref{em2}) can be well approximated by $(1+\frac{m}{\hbar^2k_{\rm F}}dU_p/dk)(1+\frac{m}{\hbar^2k_{\rm F}}dU_n/dk)\approx (1+\frac{m}{\hbar^2k_{\rm F}}dU_0/dk)^2=(m/m^*_0)^2$ ~\cite{LiBA13}. Expanding the Eq.~(\ref{em2}) to the first-order in isospin asymmetry parameter $\delta$, we have
\begin{equation}\label{npe1}
m^*_{n-p}\approx 2\delta\frac{m}{\hbar^2k_F}\left[-\frac{dU_{sym,1}}{dk}-\frac{k_F}{3}\frac{d^2U_0}{dk^2}+\frac{1}{3}\frac{dU_0}{dk}\right]_{k_F}\left(\frac{m^*_0}{m}\right)^2.
\end{equation}
While the above expressions are valid at arbitrary densities, in comparing with the nucleon optical potentials from nucleon-nucleus scattering experiments, we shall apply them only at the saturation density $\rho_0$. It is interesting to note that the above equation indicates that the $m^*_{n-p}$ depends apparently on the momentum dependence of  both the isovector and isoscalar potentials. However, as we shall show numerically, the
last two terms, i.e., $-k_F/3\cdot d^2U_0/dk^2$ and $1/3\cdot dU_0/dk$, largely cancel out each other, leaving the momentum dependence of the isovector potential $dU_{sym,1}/dk$ as the dominating factor.

\subsection{ Connecting the nucleon optical model potential with its potential in isospin-asymmetric nucleonic matter}
How can one obtain the $U_0(\rho_0,k)$, $U_{sym,1}(\rho_0,k)$ and $U_{sym,2}(\rho_0,k)$ from the $\mathcal{U}_0(\mathcal{E})$, $\mathcal{U}_{sym,1}(\mathcal{E})$ and $\mathcal{U}_{sym,2}(\mathcal{E})$ extracted from optical model analyses of nucleon-nucleus scattering experiments at the beam energy $\mathcal{E}$? The answer can be found partially in Refs.~\cite{Dab64,LiXH13}. Here we summarize their relationship and supplement a few key equations necessary for conveniently transforming one to the other. Since we are only considering the transformation at normal density while the momentum $k$ and kinetic energy {T} are trivially related, we shall now use the $T_{\tau}$ and $\delta$ as two independent variables necessary in expressing the three parts of the nucleon potential given in Eq.~(\ref{sp}). According to Ref.~\cite{Dab64}, we simply have
\begin{equation}\label{Uu}
\mathcal{U}_{\tau}(\mathcal{E},\delta)=U_{\tau}(T_{\tau}(\mathcal{E}),\delta)
\end{equation}
but one has to be very careful about the different dispersion relationship $T_{\tau}(\mathcal{E})$ for neutrons and protons because of the momentum dependence of their isovector potential. In symmetric nuclear matter, the dispersion relationship $T(\mathcal{E})$ can be readily obtained from manipulating the single-nucleon energy
\begin{align}\label{Dispersion2}
\mathcal{E}=T+U_0(T)
\end{align}
once the momentum dependence of the isoscalar potential $U_0(T)$ is known.
For the same nucleon energy $\mathcal{E}$, by expanding the  $U_{\tau}(T_{\tau})$ to the first-order in $\delta$, one obtains the kinetic energy $T_{\tau}(\mathcal{E})$ for protons and neutrons in asymmetric matter in terms of the $T(\mathcal{E})$ as
\begin{align}\label{Disp3}
T_{\tau}(\mathcal{E})=T(\mathcal{E})-\tau_3U_{sym,1}(T)\mu(T)\cdot\delta
\end{align}
where $\mu=(1+dU_0/dT)^{-1}$.
Inserting the above relationship into Eq.~(\ref{sp}) and expanding all terms up to $\delta^2$, the Eq.~(\ref{Uu}) then leads to the following transformation relations~\cite{Dab64,LiXH13}
\begin{align}\label{Tran}
&U_0(T(\mathcal{E}))=\mathcal{U}_0(\mathcal{E}),\,\,\,\,\,U_{sym,1}(T(\mathcal{E}))=\frac{\mathcal{U}_{sym,1}}{\mu},\\
&U_{sym,2}(T(\mathcal{E}))=\frac{\mathcal{U}_{sym,2}}{\mu}+\frac{\zeta\mathcal{U}_{sym,1}}{\mu^2}+\frac{\vartheta\mathcal{U}^2_{sym,1}}{\mu^3},\nonumber
\end{align}
where
\begin{align}
\mu=1-\frac{\partial \mathcal{U}_0}{\partial \mathcal{E}},\,\,\,\,\zeta=\frac{\partial \mathcal{U}_{sym,1}}{\partial \mathcal{E}},\,\,\vartheta=\frac{\partial^2 \mathcal{U}_0}{\partial \mathcal{E}^2}
\end{align}

Thus, the isoscalar effective mass $m_0^*/m$ can be extracted directly using the nucleon isoscalar optical potential. To extract the neutron-proton effective mass splitting, however, the factor $\mu$ has to be included. We also notice that the Coulomb potential is explicitly considered in the optical model analyses of proton-nucleus scattering data. Moreover, we consider the theoretically uncharged isospin-asymmetric nucleonic matter without the requirement of being in $\beta$ equilibrium. Thus, the above transformations are valid for both neutrons and protons. For transformations to the interior of nuclei in $\beta$ equilibrium an extra relationship between the Coulomb potential and the symmetry potential is required~\cite{Dab64,Sat58}.

\section{ Results and Discussions}
Our work is carried out using the modified APMN code~\cite{Shen02} which has been applied extensively during the last decade in optical model analyses of various aspects of nucleon-nucleus reactions. Technical details of the code and examples from earlier analyses of some portions of the available data for other purposes can be found in Refs.~\cite{LiXH12,Shen02,LiXH07,LiXH08}.
We use totally 37 parameters in the optical model potential. To find the optimal parameter set we perform a global $\chi^2$ minimization using all available nucleon-nucleus reaction (i.e., non-elastic) and elastic angular differential cross sections below about 200 MeV from the EXFOR database~\cite{Exfor}. To check the reliability of our conclusions, we performed the following three analyses: Case I for neutron-nucleus, Case II for proton-nucleus and Case III for all nucleon-nucleus scattering.
Here we use the average $\chi^2$ per nucleus defined as
\begin{equation}
\chi^2=\frac{1}{N}\sum^N_{n=1}\chi^2_n
\end{equation}
with $\chi^2_n$ for each single nucleus $n$ calculated from
\begin{widetext}
\begin{align}
\chi^2_n=\left(\frac{W_{n,non}}{N_{n,non}}\sum^{N_{n,non}}_{i=1}\left(\frac{\sigma^{th}_{non,i}-\sigma^{exp}_{non,i}}{\Delta\sigma^{exp}_{non,i}}\right)^2
+\frac{W_{n,el}}{N_{n,el}}\sum^{N_{n,el}}_{i=1}\frac{1}{N_{n,i}}\sum^{N_{n,i}}_{j=1}\left(\frac{\sigma^{th}_{el}(i,j)-\sigma^{exp}_{el}(i,j)}{\Delta \sigma^{exp}_{el}(i,j)}\right)^2\right)/\left(W_{n,non}+W_{n,el}\right)
\end{align}
\end{widetext}
where $N$ is the total number of nuclei included in the parameter optimization. The $\sigma^{th}_{el}(i,j)$ and $\sigma^{exp}_{el}(i,j)$ are the theoretical and experimental elastic differential cross sections at the $j$th angle with the $i$th incident energy, respectively. The $\Delta\sigma^{exp}_{el}(i,j)$ is the corresponding experimental uncertainty. $N_{n,i}$ denotes the number of angles where the data are taken for the $n$th nucleus at the $i$th incident energy. $N_{n,el}$ is the number of incident energy for elastic scattering on the $n$th nucleus. The $\sigma^{th}_{non,i}$ and $\sigma^{exp}_{non,i}$ are the theoretical and experimental non-elastic (reaction) cross sections at the $i$th incident energy, respectively. The $\Delta\sigma^{exp}_{non,i}$ is
the corresponding experimental uncertainty. While the $N_{n,non}$ is the number of nonelastic cross sections available for the $n$th nucleus. The $W_{n,el}$ and $W_{n,non}$ are the weighting factors of the elastic angular differential and nonelastic cross sections, respectively. They are chosen according to the numbers of the respective experimental data available. For the Case I, only the elastic differential cross sections are used, for the Case II both the nonelastic and elastic differential cross sections are used while the Case III is a simultaneous analysis of all data considered in the Case I and II.

\begin{table}[!htb]
\caption{The values and the corresponding standard deviation (error bar) for the
parameter $V_i$ ($i=\mathrm{0,1,2,3,3L,4,4L}$) obtained from 1161 sets of  neutron-nucleus scattering experimental data involving 104 targets.}
\begin{tabular}{ccc}
\hline\label{Tab1}
  parameter & average value & error bar \\
  \hline
$V_0(\mathrm{MeV})$       &54.96                   &1.13    \\
$V_1$                     &$-$0.3391                 &0.0211   \\
$V_2(\mathrm{MeV}^{-1})$  &2.312$\times$$10^{-4}$   &1.243$\times$$10^{-4}$\\
$V_3(\mathrm{MeV})$       &$-$25.43                 &6.13\\
$V_{\mathrm{3L}}$         &0.2062                   &0.0487 \\
$V_4(\mathrm{MeV})$       &$-$8.832                 &4.541\\
$V_{\mathrm{4L}}$         &3.931$\times$$10^{-4}$   &9.252 $\times$$10^{-4}$\\
\hline \hline
\end{tabular}
\end{table}
\begin{table}[!htb]
\caption{The values and the corresponding standard deviation (error bar) for the
parameter $V_i$ ($i=\mathrm{0,1,2,3,3L,4,4L}$) obtained from 1088 sets of  proton-nucleus scattering experimental data involving 130 targets.}
\begin{tabular}{ccc}
\hline\label{Tab2}
  parameter & average value & error bar \\
  \hline
$V_0(\mathrm{MeV})$       &54.93                   &1.03    \\
$V_1$                     &$-$0.3242                 &0.0311  \\
$V_2(\mathrm{MeV}^{-1})$  &2.433$\times$$10^{-4}$   &1.152$\times$$10^{-4}$\\
$V_3(\mathrm{MeV})$       &$-$24.94                &5.98 \\
$V_{\mathrm{3L}}$         &0.2151                    &0.0552 \\
$V_4(\mathrm{MeV})$       &$-$8.647                 &4.315\\
$V_{\mathrm{4L}}$         &3.642$\times$$10^{-4}$   &8.623 $\times$$10^{-4}$\\
\hline \hline
\end{tabular}
\end{table}
\begin{table}[!htb]
\caption{The values and the corresponding standard deviation (error bar) for the
parameter $V_i$ ($i=\mathrm{0,1,2,3,3L,4,4L}$) obtained using all nucleon-nucleus scattering experimental data involving 234 targets.}
\begin{tabular}{ccc}
\hline\label{Tab3}
  parameter & average value & error bar \\
  \hline
$V_0(\mathrm{MeV})$       &55.06                    &1.24   \\
$V_1$                     &$-$0.3432                 &0.0304  \\
$V_2(\mathrm{MeV}^{-1})$  &2.524$\times$$10^{-4}$   &1.224$\times$$10^{-4}$\\
$V_3(\mathrm{MeV})$       &$-$25.40                 &6.27\\
$V_{\mathrm{3L}}$         &0.2051                    &0.0562\\
$V_4(\mathrm{MeV})$       &$-$8.896                 &4.864\\
$V_{\mathrm{4L}}$         &3.844$\times$$10^{-4}$   &10.721 $\times$$10^{-4}$\\
\hline \hline
\end{tabular}
\end{table}
We note here that in the past the uncertainties of the optical model parameters are normally estimated by dividing randomly the considered data sets into two equal parts and then evaluating the resulting differences in the model parameters. In the present work, we carry out a covariance analysis ~\cite{Rei10,Fat11} of the model parameters around their optimal values by analyzing the error matrix using the complete data set. The standard deviations of all model parameters are then evaluated consistently and uniformly. The minimum (total instead of per degree of freedom) $\chi^2$ are 50.62, 54.75 and 65.69, respectively, for the three cases studied. The most relevant parameters and their variances for the purpose of this work are summarized in Tables~\ref{Tab1},~\ref{Tab2} and~\ref{Tab3}, respectively.  It is worth noting that the errors considered in this work are all statistical in nature.
Systematic errors are also important but hard to estimate. We admit here that no systematic error due to the model assumptions, such as the shape of the optical potential, has been studied yet in this work.

As an illustration of the quality of the global fit to the experimental data, shown in Fig.\  1 is a typical example of the angular differential cross sections for $n+^{208}$Pb (left) and $p+^{208}$Pb (right) reactions. For a comparison, shown also in Fig.\ 1 are the plot using optical model parameters given by Koning {\it et al.}~\cite{Kon03} from analyzing nucleon-nucleus scattering data. It is seen that both ours and the Koning parameterization describe the data reasonably well. More quantitatively, the Koning parameters lead to $\chi^2$ values of 48.35 and 50.85, respectively, for the neutron-nucleus and proton-nucleus scattering. They are both compatible with ours.
\begin{figure}
\centering
\includegraphics[width=9cm]{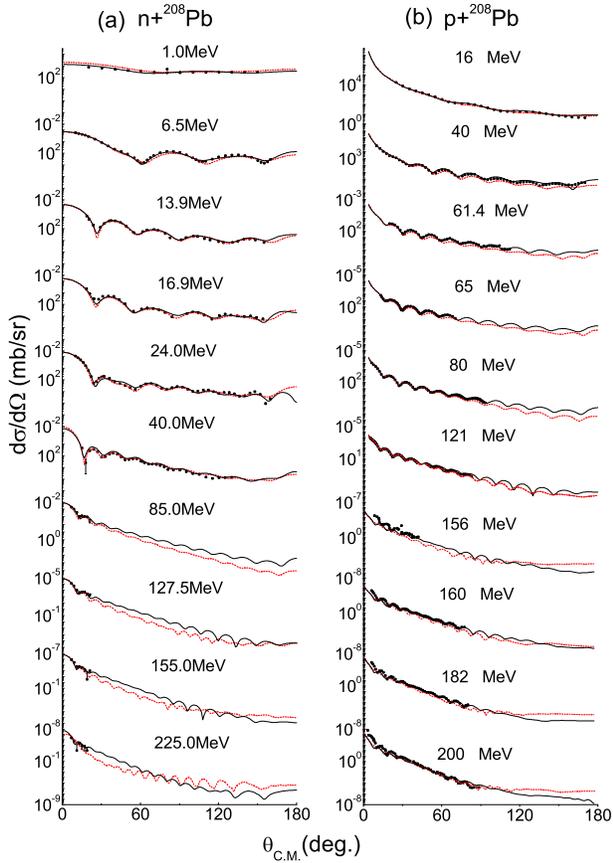}
\caption{(Color online) Angular differential cross sections for $n+^{208}$Pb (left)
and $p+^{208}$Pb scattering (right). The dots are the experimental data, the red curves are our  calculations while the back curves are the results of Ref. ~\cite{Kon03}}
\label{fig1}
\end{figure}

\begin{figure}
\centering
\includegraphics[width=9cm]{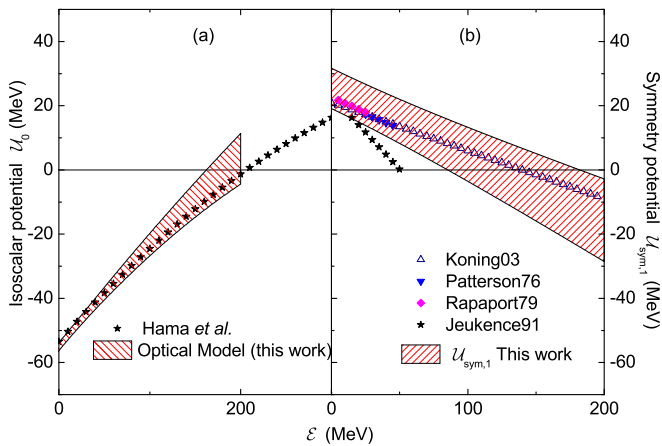}
\caption{(Color online) Energy dependent isoscalar $\mathcal{U}_0$(left) and isovector
$\mathcal{U}_{sym,1}$ (right) potentials from the present work (hatched bands) in comparison with the Schr$\ddot{\mathrm{o}}$dinger equivalent isoscalar potential obtained by Hama \textit{et al.}~\cite{Ham90} and several parameterizations for the $\mathcal{U}_{sym,1}$ from earlier studies~\cite{Kon03,Jeu91,Rap79,Pat76}.}
\label{fig2}
\end{figure}
\begin{figure}
\centering
\includegraphics[width=9cm]{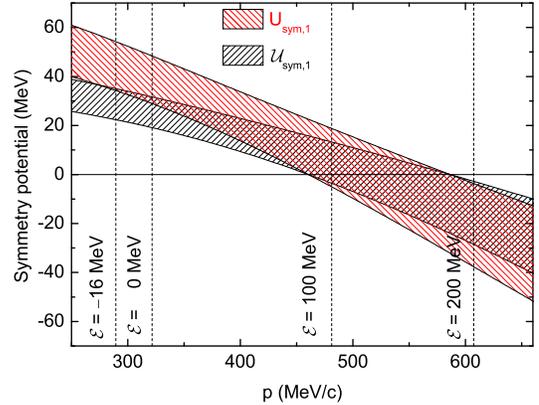}
\caption{(Color online) Momentum dependence of the symmetry potential in the nucleon optical potential $\mathcal{U}_{sym,1}$ (black) and nuclear matter $U_{sym,1}$ (red), respectively.}
\label{fig3}
\end{figure}
With the optimized optical model parameters we can then evaluate the energy/momentum dependence of both the isoscalar and isovector potentials. For this purpose, we shall use the
parameters in Table III from the simultaneous analyses of all nucleon-nucleus scattering data.
Shown in Fig.\ 2 are the $\mathcal{U}_0$ (left) and $\mathcal{U}_{sym,1}$ (right) from the present work (hatched bands) together with the Schr$\ddot{\mathrm{o}}$dinger equivalent isoscalar potential obtained by Hama \textit{et al.}~\cite{Ham90} and several parameterizations for the $\mathcal{U}_{sym,1}$ from earlier studies~\cite{Kon03,Jeu91,Rap79,Pat76}. It is seen clearly that our isoscalar potential is in good agreement with that from the Dirac phenomenology in the energy range considered. Earlier parameterizations for the $\mathcal{U}_{sym,1}$ are valid in different energy ranges. The one by Koning \textit{et al.}~\cite{Kon03} is valid up to 200 MeV as in our analyses. While others are mostly for low energies, for example, the one by Rapaport \textit{et al.}~\cite{Rap79} is for energies from 7 to 26 MeV.  It is interesting to see that our $\mathcal{U}_{sym,1}$ is consistent with earlier results, except the one by Jeukenne \textit{et al.}~\cite{Jeu91}, within our error bands.

Shown in Fig.\ 3 is a comparison of the symmetry potential $\mathcal{U}_{sym,1}$ in the nucleon optical potential and the $U_{sym,1}$  in isospin-asymmetric nucleonic matter as a function of nucleon momentum. It is seen that their slopes are significantly different especially around the nucleon Fermi momentum of 270 MeV/c. We emphasize that the momentum dependence of the $U_{sym,1}$ at normal density obtained here provides a significant boundary condition for the isovector potentials used in transport model simulations of heavy-ion reactions especially those induced by rare isotopes~\cite{LiBA08}. The $U_{sym,1}$ instead of the  $\mathcal{U}_{sym,1}$ should be used to evaluate the neutron-proton effective mass splitting.

We now turn to the evaluation of both the nucleon isoscalar effective mass $m^{*}_0/m$
and the neutron-proton effective mass splitting $m^{*}_{n-p}$. Using the Eqs.~(\ref{em1}),~(\ref{npe1}) and~(\ref{Tran}), they can readily be expressed in terms of the optical model potential parameters as
\begin{equation}
\frac{m^{*}_0}{m}=\left[1+V_1+2V_2\mathcal{E}\right]_{k_F},
\end{equation}
and
\begin{widetext}
\begin{equation}
m^*_{n-p}(\rho_0,\delta)= 2\delta\cdot\left[V_{3L}-\frac{2V_2(V_3+V_\mathrm{3L}\mathcal{E})}{1+V_1+2V_2\mathcal{E}}-\frac{2}{3}\frac{\hbar^2k_F^2}{m}
\frac{V_2(V_1+2V_2\mathcal{E})}{1+V_1+2V_2\mathcal{E}}+\frac{2}{3}\frac{\hbar^2k_F^2}{m}V_2\right]_{k_F}.
\end{equation}
\end{widetext}
Choosing the single-nucleon energy at normal density $\rho_0$ to be $\mathcal{E}_0=-16$ MeV
(where $k=k_F$), and using the values and corresponding errors for the $V_i(i=1,2,3,3\mathrm{L})$ given in Tables I, II and III, we obtain the results shown in Table IV.

\begin{table}[!htb]
\caption{Nucleon isoscalar effective mass $m^{*}_0/m$ and the neutron-proton effective mass splitting $m^{*}_{n-p}$ from the three cases studied in this work.}
\begin{tabular}{ccc}
\hline\label{Tab4}
Case   & $m^{*}_0/m$ & $m^{*}_{n-p}(\delta)$ \\
  \hline
I  &$0.65\pm 0.05$ & $0.41\pm0.14$ \\
II & $0.67\pm 0.06$ & $0.44\pm0.16$ \\
III & $0.65\pm 0.06$ &$0.41\pm0.15$\\
\hline \hline
\end{tabular}
\end{table}
\begin{table}[!htb]
\caption{Sources of the neutron-proton effective mass splitting $m^{*}_{n-p}$ at normal density.}
\begin{tabular}{cccc}
\hline\label{Tab5}
Case & $-dU_{sym,1}/dk$ & $-k_F/3d^2U_0/dk^2$ & $1/3dU_0/dk$ \\
  \hline
I  & 29.39  & -12.07 & 9.74 \\
II & 29.76  & -11.68 & 9.33 \\
III & 30.44 & -12.83 & 10.17 \\
\hline \hline
\end{tabular}
\end{table}
The results from the three cases are consistent within the error bars. We notice that the isoscalar effective mass extracted here is consistent with the empirical values from many other analyses, see, e.g.,  the often quoted value of $m^{*}_0/m=0.70\pm 0.05$ from Refs.~\cite{jamo,Jeu76}. 
As usual, the resulting isocalar effective mass from the optical model potential is less than the theoretical prediction within the Brueckner-Hartree-Fock
approach \cite{Jeu76} that is typically closer to unity. This feature was understood as the local enhancement of the nucleon effective mass at the Fermi surface due to
the core polarized states with low excitation energy \cite{Jeu76,Ber68} and  which are not included in the optical model analysis. The values of the neutron-proton effective mass splitting $m^{*}_{n-p}$ extracted here are appreciably larger than the earlier value of $(m_n^*-m_p^*)/m=(0.32\pm 0.15) \delta$ extracted directly from taking an average of the available nucleon isovector optical potentials~\cite{XuC10} without performing the transformation discussed earlier. However, they overlap largely within the statistical error bars. While the current uncertainty of about $37\%$ is not fundamentally better than the previous one,  the present analysis is much more meaningful due to the method and the large number of independent data sets used directly. To our best knowledge, the value of $m^{*}_{n-p}=(0.41\pm0.15)\delta$ extracted in the Case III is presently the most reliable and stringent constraint on the neutron-proton effective mass splitting in isospin asymmetric nucleonic matter at normal density. For neutron-rich matter, the effective mass of neutrons is definitely larger than that of protons. This finding is consistent with many model predictions, see, e.g., \cite{LLL, ZHL}, but disagrees with many others. 

As we noticed earlier in Eq.~(\ref{npe1}), the $m^{*}_{n-p}$ comes from the momentum dependence of both the isovector and isoscalar potentials. What are their respective contributions? To answer this  question, summarized in Table V are the values of $-dU_{sym,1}/dk$, $-k_F/3d^2U_0/dk^2$, and $1/3dU_0/dk$ at $k_F$ extracted from the data. It is seen that the last two terms due to the momentum dependence of the isoscalar potential largely cancel out, leaving the momentum dependence of the isovector potential $-dU_{sym,1}/dk$ as the dominating source of the neutron-proton effective mass splitting $m^{*}_{n-p}$ at normal density.

\section{Summary}
In summary,  within an isospin dependent optical potential model using all existing data of nucleon-nucleus
reaction and elastic angular differential cross sections up to about 200 MeV, we extracted the momentum dependence of both the nucleon isoscalar and isovector potentials at normal density. The isoscalar potential is consistent with the Hama potential from earlier analyses using a relativistic optical potential mmodel. The extracted potentials can be used to calibrate the isospin-dependent nucleon potentials used in transport model simulations of nuclear reactions and  provide a useful boundary condition to test predictions by various nuclear many-body theories.  The extracted nucleon isoscalar effective mass is consistent with its empirical values extracted earlier in the literature. Most importantly, the neutron-proton effective mass splitting is found to be $m^{*}_{n-p}=(0.41\pm0.15)\delta$. We believe it is presently the most reliable value for this very poorly known but rather important quantity for resolving many interesting issues in both nuclear physics and astrophysics.

\begin{acknowledgments}
We would like to thank Chong-Hai Cai and Bao-Jun Cai for useful discussions. This work was supported in part by the US National Science Foundation under Gront No. PHY-1068022, National Aeronautics and Space Administration under Grant NNX11AC41G issued through the Science Mission Directorate and the CUSTIPEN (China-U.S. Theory Institute for Physics with Exotic Nuclei) under DOE Grant No. DE-FG02-13ER42025; NNSF of China under Grant Nos.11047157, 11205083, 11275125 and 11135011, 11320101004, the Shanghai Rising-Star Program under grant No. 11QH1401100, the  Shu Guang  project supported by Shanghai Municipal Education Commission and Shanghai Education Development Foundation, the Program for Professor of Special Appointment (Eastern Scholar) at Shanghai Institutions of Higher Learning, and the Science and Technology Commission of Shanghai Municipality (11DZ2260700).
\end{acknowledgments}


\begin{references}

\bibitem{sjo76} O. Sj\"oberg, Nucl. Phys. A {\bf 265}, 511 (1976).

\bibitem{Bom91} I. Bombaci, U. Lombardo, Phys. Rev. C \textbf{44}, 1892 (1991).

\bibitem{Dob99} J. Dobaczewski, Acta Phys. B \textbf{30}, 1647 (1999).

\bibitem{LiBA08} B.A. Li, L.W. Chen, C. M. Ko, Phys. Rep. \textbf{464}, 113 (2008).

\bibitem{zuo05} W. Zuo, L.G. Cao, B.A. Li, U. Lombardo and C.W. Shen, Phys. Rev. C {\bf 72}, 014005 (2005).

\bibitem{Fuchs1} E. N. E. van Dalen, C. Fuchs and Amand Faessler, Phys. Rev. C {\bf 72}, 065803 (2005).

\bibitem{Fuchs2} C. Fuchs and H.H. Wolter, Eur. Phys. J. A {\bf 30}, 5 (2006).

\bibitem{Bom01} I. Bombaci, Chap. 2 in Isospin Physics in Heavy-Ion Col- lisions at Intermediate Energies, Eds. B. A. Li and W. Udo Schr\"oder (Nova Science Publishers, Inc, New York, 2001).

\bibitem{Riz04} J. Rizzo, M. Colonna, M, Ditoro, V. Greco, Nucl. Phys. A \textbf{732}, 202 (2004).

\bibitem{Li04} B.A. Li, Phys. Rev. C {\bf 69}, 064602 (2004).

\bibitem{OuLi11} Li Ou, Zhuxia Li, Yingxun Zhang, Min Liu, Phys. Lett. B {\bf 697}, 246 (2011).

\bibitem{XuC10} C. Xu, B.A. Li, and L.W. Chen, Phys. Rev. C \textbf{82}, 054607 (2010).

\bibitem{XuC11} C. Xu, B.A. Li, L.W. Chen, and C.M. Ko, Nucl. Phys. A \textbf{865}, 1 (2011).

\bibitem{Rchen} R. Chen, B.J. Cai, L.W. Chen, B. A. Li, X.H.  Li and C. Xu, Phys. Rev. C  {\bf 85}, 024305 (2012).

\bibitem{hug} N.M. Hugenholtz and L. van Hove, Physica \textbf{24}, 363 (1958).

\bibitem{TEsym} B.A. Li, A. Ramos, G. Verde, and I. Vida\~na, eds., ``Top-
ical issue on nuclear symmetry energy", Eur. Phys. J. A {\bf 50}, No. 2, (2014).

\bibitem{LiBA13} B.A. Li, X. Han, Phys. Lett. B \textbf{727}, 276 (2013).

\bibitem{LiBA04} B.A. Li, C.B.Das, S. Das Gupta, C.Gale, Phys. Rev. C {\bf 69}, 011603 (2004);
{\it ibid}, Nucl. Phys. A {\bf 735},563 (2004).

\bibitem{Riz05} J. Rizzo, M.Colonna and M. Di Toro, Phys. Rev. C {\bf 72}, 064609 (2005).

\bibitem{Feng12} Zhao-Qing Feng, Nucl. Phys. A {\bf 878}, 3(2012) and Phys. Lett. B {\bf 707},83(2012).

\bibitem{Zhang14} Yingxun Zhang, M. B. Tsang, Zhuxia Li and Hang Liu,  Phys. Lett. B {\bf 732}, 186 (2014).

\bibitem{Exfor}\url{http://www.nndc.bnl.gov/}

\bibitem{Rei10} P.-G. Reinhard and W. Nazarewicz, Phys. Rev. C {\bf 81}, 051303 (2010).

\bibitem{Fat11} F.J. Fattoyev, J. Piekarewicz, Phys. Rev. C \textbf{84}, 064302 (2011).

\bibitem{Hod94} P.E. Hodgson, The Nucleon Optical Model, 1994 (World Scientific).

\bibitem{Yon98} P.G. Yong, RIPL Handbook, Vol. 41, 1998,
http://www-nds.iaea.org/ripl/, Ch.4: Optical Model Parameters.

\bibitem{Var91} R.L. Varner, W.J. Thompson, T.L. Mcabee, E.J. Ludwig, and T.B. Clegg, Phys. Rep. \textbf{201}, 57 (1991).

\bibitem{Kon03} A.J. Koning and J.P. Delaroche, Nucl. Phys. A \textbf{713}, 231 (2003).

\bibitem{Wep09} S.P. Weppner, R.B. Penney, G.W. Diffendale, and G. Vittorini, Phys. Rev. C \textbf{80}, 034608 (2009).

\bibitem{Han10} Y.L. Han, Y.L. Xu, H.Y. Liang, H.R. Guo, and Q.B. Shen, Phys. Rev. C \textbf{81}, 024616 (2010).

\bibitem{LiXH12} X.H. Li, L.W. Chen, Nucl. Phys. A  \textbf{874}, 62 (2012).

\bibitem{AnHX06} H.X. An, C.H. Cai, Phys. Rev. C  \textbf{73}, 054605 (2006).

\bibitem{Lan62} A.M. Lane, Nucl. Phys. \textbf{35}, 676 (1962).

\bibitem{jamo} M. Jaminon, C. Mahaux, Phys. Rev. C \textbf{40}, 354 (1989).

\bibitem{LLL}L.L. Li, Z.H. Li, E.G. Zhao, S.G. Zhou, W. Zuo, A. Bonaccorso and U. Lombardo, Phys. Rev. C {\bf 80}, 064607 (2009).

\bibitem{Dab64} J. Dabrowski, Phys. Lett. \textbf{8}, 90 (1964).

\bibitem{LiXH13} X.H. Li, B.J. Cai, L.W. Chen, R. Chen, B.A. Li, C. Xu, Phys. Lett. B  \textbf{721}, 101 (2013).

\bibitem{Sat58}G.R. Satchler, Phys. Rev. {\bf 109}, 429 (1958).

\bibitem{Shen02} Q.B. Shen, Nucl. Sci. Eng. \textbf{141}, 78 (2002).

\bibitem{LiXH07} X.H. Li, C.T. Liang, C.C. Hai, Nucl. Phys. A \textbf{789}, 103 (2007).

\bibitem{LiXH08} X.H. Li, C.H. Cai, Nucl. Phys. A  \textbf{801}, 43 (2008).

\bibitem{Ham90} S. Hama, B.C. Clark, E.D. Cooper, H.S. Sherif, R.L. Mercer, Phys. Rev. C \textbf{41}, 2737 (1990).

\bibitem{Jeu91} J.-P. Jeukenne, C. Mahaux, R. Sartor, Phys. Rev. C \textbf{43}, 2211 (1991).

\bibitem{Rap79} J. Rapaport, V. Kulkarni, R.W. Finlay, Nucl. Phys. A \textbf{330}, 15 (1979).

\bibitem{Pat76} D.M. Patterson, R.R. Doering, A. Galonsky, Nucl. Phys. A \textbf{263}, 261 (1976).

\bibitem{Jeu76} J.P. Jeukenne, A. Lejeune, C. Mahaux, Phys. Rep. \textbf{25}, 83 (1976).

\bibitem{Ber68} G.F. Bertsch and T.T.S. Kuo, Nucl. Phys. A112 (1968) 204.

\bibitem{ZHL}Z.H. Li and U. Lombardo, Phys. Rev. C{\bf 78}, 047603 (2008).

\end{references}
\end{document}